\documentclass[aps,pra,groupedaddress,superscriptaddress,twocolumn,showpacs]{revtex4-1}

\usepackage[utf8x]{inputenc}
\usepackage{color}
\usepackage{bbm} 

\usepackage{amsfonts,amsmath,amssymb,stmaryrd}

\usepackage{graphicx}

\usepackage{subfigure}  

\usepackage{bbm} 

\usepackage{hyperref}

\usepackage{bbm}

\usepackage{epsfig}

\usepackage{mathrsfs}

\usepackage{verbatim}

\usepackage{ulem}	

\renewcommand{\l}{\left(}
\renewcommand{\r}{\right)}
\newcommand{\bra}[1]{\langle#1|}
\newcommand{\ket}[1]{|#1\rangle}

\newcommand{\LN}{\text{LN}}
\newcommand{\qh}{\text{qh}}
\newcommand{\eff}{\text{eff}}

\renewcommand{\H}{\hat{\mathcal{H}}}

\renewcommand{\a}{\hat{a}}

\newcommand{\ad}{\hat{a}^\dagger}
\newcommand{\bd}{\hat{b}^\dagger}

\newcommand{\hc}{\text{h.c.}}

\usepackage{array}

\usepackage{cancel,ifthen}

\newcommand{\Komment}[2][NoInPuT]{\ifthenelse{\equal{#1}{NoInPuT}}{}{{\color{red}\sout{#1}}} {\color{blue} #2}}

\usepackage{bm}	

\begin{document}
\normalem	

\title{Topological growing of Laughlin states in synthetic gauge fields}

\author{Fabian Grusdt}
\affiliation{Department of Physics and Research Center OPTIMAS, University of Kaiserslautern, Germany}
\affiliation{Graduate School Materials Science in Mainz, Gottlieb-Daimler-Strasse 47, 67663 Kaiserslautern, Germany}

\author{Fabian Letscher}
\affiliation{Department of Physics and Research Center OPTIMAS, University of Kaiserslautern, Germany}

\author{Mohammad Hafezi}
\affiliation{Joint Quantum Institute, NIST/University of Maryland, College Park MD}
\affiliation{ECE Department and Institute for Research in Electronics and Applied Physics, University of Maryland, College Park, MD 20742, USA}

\author{Michael Fleischhauer}
\affiliation{Department of Physics and Research Center OPTIMAS, University of Kaiserslautern, Germany}



\begin{abstract}
We suggest a scheme for the preparation of highly correlated Laughlin (LN) states in the
presence of synthetic gauge fields, realizing an analogue of the fractional quantum Hall effect
in photonic or atomic systems of interacting bosons. It is based on the idea of growing such states
by adding weakly interacting composite fermions (CF) along with magnetic flux quanta one-by-one. 
The topologically protected Thouless pump ("Laughlin's argument") is used to create two localized flux quanta 
and the resulting hole excitation is subsequently filled by a single boson, which, together with
one of the flux quanta forms a CF. Using our protocol, filling $1/2$ LN states can be grown with particle number $N$ increasing linearly in time and strongly suppressed
number fluctuations. To demonstrate the feasibility of our scheme, we consider two-dimensional (2D) lattices 
subject to effective magnetic fields and strong on-site interactions. 
We present numerical simulations of small lattice systems and discuss also the influence of losses.
\end{abstract}

\date{\today}

\maketitle

\paragraph*{Introduction}
In recent years topological states of matter 
\cite{Thouless1982,Laughlin1983,Halperin1984,AROVAS1984,MOORE1991,Wen1995,
Bonderson2006,Kitaev2006a} have attracted a great deal of interest, partly due to 
their astonishing physical properties (like fractional charge and statistics) but 
also because of their potential practical relevance for quantum 
computation \cite{Kitaev2003,Nayak2008}. While these exotic phases of matter were 
first explored in the context of the quantum Hall effect of electrons subject to 
strong magnetic fields \cite{Vonklitzing1980,Tsui1982}, there has been considerable
progress recently towards their realization in cold-atom
\cite{Lin2009,Aidelsburger2011,Aidelsburger2013,Miyake2013} as well as photonic 
\cite{Wang2009,Hafezi2011,Kraus2012,Khanikaev2013,Rechtsman2013,Rechtsman2013a,
Hafezi2013a} systems. A particularly attractive feature of such
quantum Hall simulators are the comparatively large 
intrinsic length scales which allow coherent preparation, manipulation 
and spatially resolved detection
of exotic many-body phases and their excitations. 

In electronic systems the preparation of topological states of matter relies on quick thermalization and cooling 
below the many-body gap. While this is already hard to achieve in cold-atom systems (partly due to the 
small required temperatures), cooling is even less of an option in photonic systems due to the 
absence of effective thermalization mechanisms. On the other hand, lasers with narrow linewidths 
allow for a completely different avenue towards preparation of  extremely pure quantum states. 
For instance, it was suggested to use the coherence properties of lasers to directly excite two 
(and more) photon LN states in non-linear cavity arrays \cite{Umucalilar2012}, where the laser plays
the role of a coherent pump. However, this approach has the inherent problem of an extremely small multi-photon 
transition amplitude. While this might be acceptable for small systems of $N=2,3$ photons, it 
makes the preparation of true many-body states with $N \gg 2$ practically impossible. Moreover, the prepared
states in this case contain superpositions of different photon-numbers rather than being Fock states.

\begin{figure}[b]
\centering
\epsfig{file=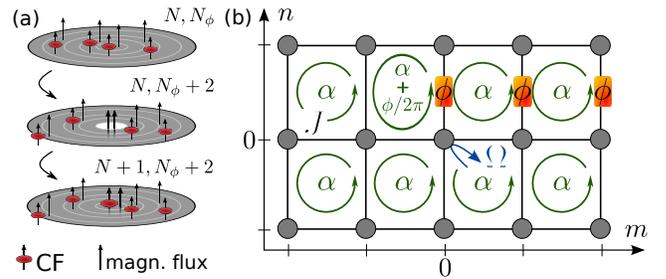, width=0.47\textwidth}
\caption{(Color online) (a) The key idea of our scheme is to grow LN states by 
introducing weakly interacting CFs into the system. This is achieved by adding 
magnetic flux (arrows) in the center and replenishing the arising hole by a new 
boson (red bullet). (b) We consider the Hofstadter-Hubbard model (flux $\alpha$ per 
plaquette). Additional flux $\phi$ can be introduced in the center
by adiabatically changing the complex phase of the hoppings marked with a box. 
Furthermore, the central site is assumed to be externally accessible for a 
coherent drive (Rabi frequency $\Omega$).}
\label{fig:setup}
\end{figure}

In this letter we suggest an alternative scheme for the preparation of topologically 
ordered states of strongly interacting bosons, specifically for the $1/2$ LN state, and we discuss 
systems allowing for an implementation of the scheme with state-of-the-art technology. 
It consists of growing such states and makes direct use of the 
Thouless pump \cite{Thouless1983} connected to the many-body topological invariant. 
In the case of quantum Hall physics the latter is realized by local 
flux insertion in the spirit of Laughlin's argument for the quantization of the Hall 
conductivity $\sigma_{H}$ \cite{Laughlin1981}: Introducing magnetic flux $\phi/2 
\pi =2$ (in units of the flux quantum) in the center of the system produces a quantized outwards Hall current $\sim 
\sigma_{H} \partial_t \phi$, leaving behind a hole, see FIG. \ref{fig:setup} (a). 

In the next step, the so-created hole can be replenished by a single boson. In view of the composite fermion (CF) picture
\cite{JAIN1989,Jain2007} of the fractional quantum Hall effect, this refilling step can be interpreted as the 
addition of a single CF (composed of a bare boson and one flux quantum) into a free orbital 
of the CF Landau level (LL), using up the remaining flux quantum. 
To refill the hole deterministically by a single boson, we consider a coherent pump in the center of the system. 
Excitations by more than one particle are prohibited by the many-body gap, and the coherent coupling can 
not decrease the total particle number because the central cavity is empty initially. Thus, our final state has sub-poissonian boson 
number statistics. A complementary scheme, where holes resulting from boson losses are dynamically 
refilled in the entire system using single photon pumps, has recently been suggested for photonic systems 
\cite{Kapit2014}. Our protocol, in contrast, does not rely on an explicit single photon source. 

A key advantage of our scheme, compared to \cite{Popp2004,Sorensen2005,Umucalilar2012}, is the ability to grow LN states with a size increasing
linearly in time. To reach $N$ particles with given fidelity $1-\varepsilon$, the 
protocol has to be carried out sufficiently slow to avoid errors in the repumping protocol. For  $\varepsilon \ll 1$ the
total required time scales like 
\begin{equation}
T \sim \frac{N^{3/2}}{\Delta_{\LN} ~ \varepsilon^{1/2}},
\label{eq:TNtarget}
\end{equation}
where $\Delta_\LN$ is the bulk many-body gap. In contrast to previously proposed schemes \cite{Popp2004,Sorensen2005,Umucalilar2012}, $T$ only grows algebraically with $N$.

\paragraph*{Model}
We consider a 2D lattice with complex hopping elements (amplitude $J$) realizing an effective magnetic 
field, supplemented by Hubbard-type on-site interactions (strength $U$). This model is illustrated in 
FIG.\ref{fig:setup} and can be described by the following Hamiltonian,
\begin{multline*}
\H_{\text{int}} + \H_0 =  \frac{U}{2} \sum_{m,n} \ad_{m,n} \a_{m,n} \l \ad_{m,n} \a_{m,n}  - 1 \r  
\\ - J \sum_{m,n} \left[ e^{- i 2 \pi \alpha n} \ad_{m+1,n} \a_{m,n} + \ad_{m,n+1} \a_{m,n} + \hc \right],
\end{multline*}
where we used Landau gauge and set $\hbar =1$. Following Jaksch and Zoller's proposal for the creation of synthetic 
gauge fields \cite{Jaksch2003}, there have been numerous suggestions how this Hamiltonian 
can be implemented in photonic \cite{Cho2008,Hafezi2011,Umucalilar2011,Umucalilar2012,Hafezi2013a}, 
circuit-QED \cite{Koch2010,Kapit2013,Hafezi2013c} or atomic \cite{Sorensen2005,Hafezi2007} 
systems, and in the last case this goal has already been achieved \cite{Aidelsburger2013,Miyake2013}.

Local flux insertion can most easily be realized by changing the hopping elements from site $(m \geq 0, n=0)$ to $(m,1)$ 
by a factor $e^{i \phi}$, see FIG.\ref{fig:setup} (b). These links are thus described by
\begin{equation}
\H_\phi = - J \sum_{m\geq0} \left[ e^{-i \phi} \ad_{m,1} \a_{m,0} + \hc \right],
\label{eq:fluxInsertion}
\end{equation}
modifying the total magnetic flux through the central plaquette to $\alpha - 
\phi/ 2 \pi$. $\H_\phi$ is motivated by recent experiments with 
photons \cite{Hafezi2011,Hafezi2013a}, where the hopping-phases can locally and temporally be 
manipulated \cite{Hafezi2013}. Finally to replenish the system with bosons, we 
place a weak coherent pump ($\Omega \ll 4 \pi \alpha J$) in the center,
\begin{equation}
\H_\Omega = \Omega e^{- i \omega t} \ad_{0,0} + \hc.
\label{eq:Hpump}
\end{equation}
In the following we present the details of our scheme, neglecting local boson losses (rate $\gamma$) for the moment. We include losses again afterwards in the discussion of the performance of our scheme.

\paragraph*{Protocol -- continuum}
We begin by discussing the continuum case when the magnetic flux per plaquette 
$\alpha \ll 1$ is small, allowing us to make use of angular momentum 
$L_z$ as a conserved quantum number. The continuum can be described by LLs, which are eigenstates of $\H_0$ in the limit $\alpha \to 0$ with energies $E_n=(n+1/2) \omega_c$  ($n=0,1,2,...$) and $\omega_c=4 \pi \alpha J$ denoting the cyclotron frequency, see e.g.\cite{Jain2007}. The magnetic length is defined as $\ell_B = a/\sqrt{2 \pi \alpha}$, where $a$ denotes the lattice constant. In symmetric gauge the single particle states of the lowest LL (LLL) are labeled by their angular momentum quantum number $l=0,1,2,...$ \cite{Jain2007} and we define boson creation operators of these orbitals as $\bd_l$. Now we discuss the preparation of filling $\nu=N/N_\phi=1/2$ LN states, but the generalization to other fillings is straightforward. 

To create the first excitation from vacuum $\ket{0}$, we switch on the coherent pump \eqref{eq:Hpump} with frequency $\omega = \omega_c/2$, which due the blockade \cite{Birnbaum2005} (caused by strong boson-boson interactions) only allows a single particle to enter the system. Since we drive locally in the center, no angular momentum is transferred and we thus arrive at the state $\ket{\Psi_1} = \bd_0 \ket{0}$. This argument is true when excitations of higher LLs can be neglected, allowing us to project the coherent pump \eqref{eq:Hpump} into the LLL, $\H_\Omega \approx  [ \bd_0 e^{- i \omega t}  \Omega_\eff^{(1)}   + \hc ] $ with $ \Omega_\eff^{(1)} = \Omega \sqrt{\alpha}$. To this end we require a weak pump, $\Omega \ll \omega_\text{c}$, sufficiently feeble also for the blockade to work, i.e. $\Omega_\eff^{(1)} \ll \Delta_{\text{LN}}$. In the continuum the gap can be estimated from $\Delta_\LN \approx \min \l V_0 , \omega_c \r$, where $V_0=U \alpha/2$ is given 
by Haldane's zeroth-order pseudo-potential \cite{HALDANE1985}. To prepare $\ket{\Psi_1}$ from $\ket{0}$ as described, the coherent pump has to be switched on for a time $T_\pi=\pi/2 \Omega^{(1)}_\eff$ (corresponding to a $\pi$-pulse in the effective two-level system defined by $\ket{0}$ and $\ket{\Psi_1}$), which works when losses are negligible, $\gamma \ll \Omega^{(1)}_\eff$ \footnote{Note that 
the coherent drive $\Omega=\kappa \sqrt{N_\Omega}$ can be related to the single-boson coupling strength $\kappa$ of
the central site to a single-mode reservoir with average boson number $N_\Omega$. To neglect additional spontaneous 
emission into the reservoir we also require $\kappa \ll \gamma$ and thus $N_\Omega \gg 1$.}.

Next, we adiabatically introduce two units of magnetic flux into the center of the system. Thereby the initial state $\ket{\Psi_1} = \bd_0 \ket{0}$ attains two units of 
angular momentum and we end up in $\ket{\Psi_2}=\bd_2 \ket{0}$ \footnote{Here, for simplicity, we assumed that the Hamiltonian \eqref{eq:fluxInsertion} for flux insertion is replaced by one with a symmetric gauge choice preserving rotational symmetry, see e.g. A. R. Kolovsky, F. Grusdt, M. Fleischhauer, Physical Review A 89, 033607 (2014).}. This state has a ring-structure with a hole in its center, which -- repeating the first step of our protocol -- can be replenished by an additional particle using the coherent pump. Because the latter only couples to the center of the system, it can not reduce the total particle number. The combined insertion of magnetic flux and a boson can be understood as addition of a single CF, with one flux-quantum binding to the boson to form a CF in the reduced magnetic field corresponding to the remaining flux quantum. Crucially, in contrast to the first step, the new state is not the simple product state $\bd_0 \ket{\Psi_2}$. Instead the blockade mechanism allows to pump only into the $N=2$ LN state $\ket{\LN,2}$, which is the only zero-energy state with the correct total angular momentum $L_z=2$, while all other states are detuned from the pumping frequency by the gap $\Delta_\LN$. As a consequence, the corresponding Rabi frequency is reduced by a Franck-Condon factor (FCF), $\Omega_\eff^{(2)}/ \Omega =\bra{\LN,2} \bd_0 \ket{\Psi_2} \sqrt{\alpha}$.

Having established our protocol for two bosons, the extension to $N$-particle LN states $\ket{\LN,N}$ is straightforward. In this case, local flux insertion is used to create  the state $\ket{2\qh,N-1}$ with two quasiholes, which are subsequently refilled by the coherent pump to prepare $\ket{\LN,N}$. The corresponding transition amplitude $\Omega_\eff^{(N)}$ is reduced by a many-body FCF, 
\begin{equation}
\Omega_\eff^{(N)}/\Omega  =\sqrt{\alpha} ~ \bra{\LN,N} \bd_0 \ket{2 \qh, N-1}.
\label{eq:OeffN}
\end{equation}
Using exact diagonalization (ED) of small systems ($N=1,...,9$) we find that 
$\Omega_\eff^{(N)}$ is nearly constant as a function of $N$ and we extrapolate $\Omega_\eff^{(\infty)} \approx  0.70  ~\Omega 
\sqrt{\alpha}$. Thus our pump works equally for large and small boson numbers. 

A natural explanation why highly correlated many-body states 
can be grown in the relatively simple fashion described above is provided by the composite fermion picture: LN states are separable (Slater 
determinant) states of non-interacting CFs filling the CF-LLL \cite{JAIN1989}. Thus, 
introducing CFs one-by-one into the orbitals of this LLL, LN states can easily be grown.

\paragraph*{Protocol -- lattice}

To ensure a sizable cyclotron gap $\omega_c$, a not too small flux per plaquette $\alpha$ 
is desirable, where lattice effects become important. We will now study this regime, which is also of great 
experimental relevance \cite{Hafezi2013a,Aidelsburger2013,Miyake2013}.
The spectrum of the Hamiltonian $\H_0(\alpha)$ is the famous Hofstadter butterfly \cite{HOFSTADTER1976}, 
consisting of a self-similar structure of magnetic sub-bands. When interactions are 
taken into account, LN-type states can still be identified at filling $\nu=1/2$
\cite{Sorensen2005,Hafezi2007}.

The basic ideas directly carry over from the continuum to the lattice case. Because the 
many-body Chern number is strictly quantized, Laughlin's argument shows that a  
hole excitation can still be created by local flux insertion. However, due to the formation 
of magnetic sub-bands, such a quasihole becomes dispersive and will propagate away from the center. 
This leaves us only a restricted time to refill the defect, and leads to a reduced efficiency 
of repumping. To circumvent this problem, we introduce a trap for quasiholes. A static, repulsive potential of the form
\begin{equation}
\H_{\text{pot}} =  \sum_{m,n} \frac{g}{\sqrt{2 \pi} \ell_B/a} e^{-(m^2+n^2) a^2 / 2 \ell_B^2}~  \ad_{m,n} \a_{m,n} 
\label{eq:Hpot}
\end{equation}
is sufficient for a gapped ground state at every point in the protocol. An alternative would be to include carefully chosen long-range hoppings leading to a completely flat band \cite{Kapit2010}.

\begin{figure}[b]
\centering
\epsfig{file=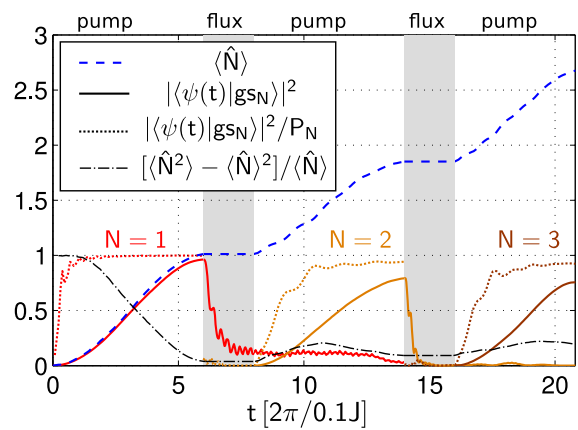, width=0.5\textwidth}
\caption{(Color online) Simulation of the full protocol on a $C_{60}$ buckyball as described in the text, for $U=10 J$ and including the static potential \eqref{eq:Hpot} with $g=J$. The overlaps (solid, conditioned on the targeted particle number $N$ - dotted) together with particle-number fluctuations (dash-dotted) indicate the accuracy of our protocol.}
\label{fig:fullProt}
\end{figure}

In the following we use ED to simulate our protocol for small systems. 
To get rid of boundary effects, which can be pronounced in small systems, we 
consider a spherical geometry \cite{Haldane1983} and take into account lattice effects 
by using a buckyball-type lattice. The hopping elements on all links have amplitude $J$ and 
their phases were chosen such that the flux per plaquette is $\alpha$. Because the 
total flux $N_\phi$ is integer quantized, it holds $\alpha=N_\phi/N_p$ with $N_p=32$ being the 
number of plaquettes. We checked numerically (using ED) that for the values of $\alpha \leq 0.2$ 
used in this paper there are gapped LN-type ground states, provided that the condition $N_\phi=2 (N-1)$ 
for $\nu=1/2$ LN states on a sphere is fulfilled. We find gaps of the order $\Delta_\LN \approx 0.1 J$, 
as predicted for a square lattice \cite{Sorensen2005,Hafezi2007}. To describe the effect of local flux insertion $N_\phi 
\rightarrow N_\phi + \phi / 2 \pi$ we slightly increase $\alpha \rightarrow \alpha + 
\phi / (2 \pi N_p)$ everywhere, except on the central plaquette where $\alpha \rightarrow \alpha 
- (1 - 1/N_p )\phi / 2 \pi$ changes by $-\phi / 2 \pi$ in thermodynamic limit (i.e. for
$N_p \rightarrow \infty$). Starting from an incompressible LN-type ground state, we checked 
numerically that the correct number of low-lying quasihole states is obtained, and that they
can be gapped out by the potential Eq.\eqref{eq:Hpot} \footnote{see Supplementary Material for more details.}.

In FIG.\ref{fig:fullProt} we present a numerical simulation of our full protocol on the $C_{60}$ buckyball lattice. We start from vacuum and $N_\phi=0$ flux quanta. Then the coherent pump Eq.\eqref{eq:Hpump} is switched on for a time $T_\Omega = 6 \pi / \Omega$ (with $\Omega = 0.05 J$) and one boson is inserted with an overlap close to one to the target $N=1$ ground state. The driving frequency $\omega$ is chosen to be resonant on the transition from the $N=0$ to the $N=1$ ground state. After introducing two more flux quanta in a time $2 \times 20 \pi/J$, of the order $2 \pi / \Delta_\LN \approx 60 / J$, the whole protocol is repeated and we finally arrive close to a three particle LN-type ground state. We find that the overlaps of the prepared states to the targeted $N$ particle ground states $\ket{\text{gs}_N}$ are close to one after all steps, and the overlaps conditioned on having the correct particle number $N$ (occurring with probability $P_N$) are even larger. At the end of the protocol, the $N=3$ boson ground state at $N_\phi=4$ is prepared with high fidelity, which carries the signatures of a LN-type state. 
Importantly the particle number fluctuations after a completed cycle are strongly suppressed  $[ \langle \hat{N}^2 \rangle-\langle \hat{N} \rangle^2]/\langle \hat{N} \rangle \ll 1$.

In our simulations we neglected edge effects and bulk losses. The latter result in a finite boson life-time, such that in the growing scheme the mean density $\rho(r)$ decays with the distance $r$ from the center. In continuum we find $\rho(r) \approx \frac{1}{4 \pi \ell_B^2} \exp \l - \gamma  T_0 \frac{r^2}{4 \ell_B^2} \r$, with $T_0$ being the duration of a single step of the protocol. In a forthcoming publication \footnote{F. Letscher, F. Grusdt,  M. Fleischhauer, in preparation} we study larger systems using a simplified model of non-interacting CFs on a lattice and show that our protocol still works when edge-effects are taken into account.

In FIG.\ref{fig:fullProt} we observe that the fidelity $\mathcal{F}_N = |\langle \psi(t) | \text{gs}_N \rangle|$ for preparation of the $N$-particle LN-type ground state is limited, mostly by the inefficiency of the pump. High fidelity, however, is a prerequisite for measuring e.g. braiding phases of elementary 
excitations, which play a central role for topological quantum computation \cite{Nayak2008}. Taking into account couplings between low-energy states of the $N$ and $N+1$ boson sectors, induced by the coherent pump \eqref{eq:Hpump}, we find the following expression for the fidelity,
\begin{equation}
\mathcal{F}_N \sim \exp \l - \l \frac{\Lambda^2}{\Delta_{\LN}^2 T_0^2} + \gamma T_0 \frac{N}{2} \r \frac{N}{2} \r.
\label{eq:scalingFidelity}
\end{equation}
The second term in the exponent describes boson loss, whereas the first term takes into account imperfections of the blockade in the repumping process with rates scaling like $\l \Omega_\eff \Lambda / \Delta_\LN \r^2$. Here $\Lambda$ is a parameter depending on non-universal FCFs, which in the continuum case $\alpha \rightarrow 0$ is found to be $\Lambda=1.4$ from finite-size extrapolations of ED results. In a lattice $\Lambda$ takes larger values and from FIG.\ref{fig:fullProt} we estimate $\Lambda \approx 10$. In Eq.\eqref{eq:scalingFidelity} we neglected fidelity losses from flux insertion, which only leads to small corrections of $\Lambda$ however, even when using the approximation $T_0 \approx T_\pi = \pi/2\Omega_\eff$. We observe a competition between losses $\sim T_0$ and errors of the pump $\sim 1/T_0^2$. Thus, for a target fidelity $\mathcal{F}_N = 1-\varepsilon$, only LN states of a restricted number of bosons $N \leq N_{\max}$ can be grown,
\begin{equation}
N_{\max} =  1.365 ~ \varepsilon^{3/5} \l \frac{\Delta_\LN}{ \Lambda \gamma} \r^{2/5}.
\end{equation}
To do so, a time $T=N_{\max} T_0=1.22~ N_{\max}^{3/2} \varepsilon^{-1/2} \Lambda / \Delta_{\LN}$ is required, which yields Eq.\eqref{eq:TNtarget}.

\paragraph*{Experimental realization}
Our protocol can be implemented in photonic cavity arrays \cite{Cho2008,Hafezi2011,Umucalilar2011,Umucalilar2012,Hafezi2013a}, where the main experimental challenges are the required large interactions $U \gtrsim J$ and small losses $\gamma \ll \Delta_\LN / N^{5/2}$. Strong non-linearities can be realized e.g. by placing single atoms into the cavities \cite{Cho2008} or coupling them to quantum dots \cite{Hafezi2013a} or Rydberg gases \cite{Peyronel2012,Grusdt2013c,Hafezi2013a}. Most promising are circuit-QED systems, where loss-rates $\gamma = \l 0.1 \text{ms}\r^{-1}$ have been achieved \cite{Devoret2013} (and $\gamma = 1 \text{ms}^{-1}$ seems feasible). The strong coupling regime can be reached and single-photon non-linearities $U=100 \text{MHz}$ are realistic \cite{Hafezi2013c}. For the case when $U \approx J$ and for $\alpha \approx 0.1$ the LN gap can be estimated to $\Delta_\LN \approx 0.05 U = 5 \text{MHz}$ \cite{Hafezi2007} which corresponds to $\Delta_\LN / \gamma \approx 3\times 10^3$. For an infidelity of $\epsilon=0.1$ this yields $N_\text{max}=7.4$ in a continuum system ($N_\text{max}=3.4$ for $\Lambda \approx 10$ as in our simulation). To observe interesting many-body physics on a qualitative level, $\epsilon=0.5$ should be sufficient which results in $N_{\text{max}}\approx 20$ in continuum. To reach even larger photon numbers, an array of multiple flux and photon pumps could be envisioned.

Alternatively, our scheme could be realized in ultra cold atomic systems \cite{Aidelsburger2013,Miyake2013}, where large interactions $U$ and negligible decay $\gamma$ are readily available \cite{Bloch2008}. In this case an idea for realizing local flux insertion would be to use optical Raman beams with non-zero angular momentum \cite{Nandi2004}, or as an alternative quasiholes could be introduced by placing a focused laser-beam close to the edge of the system and increasing its intensity adiabatically \cite{Paredes2001}. Independent of the system, means for detecting LN-type ground states are required and several approaches were discussed how this can potentially be achieved \cite{Umucalilar2012,Bhat2007,Palmer2008,Cazalilla2003,Sorensen2005,Read2003,Palmer2006}.

\paragraph*{Summary \& Outlook}
We proposed a scheme for the preparation of highly correlated LN states of bosons in artificial gauge fields.
LN states can be understood in terms of weakly interacting CFs, and our protocol is 
based on the idea of growing non-correlated states of the latter. We 
demonstrated that this can be achieved by first creating LN quasihole excitations 
which are subsequently refilled with bosons. Importantly, our protocol only requires a 
preparation time scaling slightly faster than linear with system-size.

Our scheme is not restricted to the preparation of LN states of bosons. For example, we expect that the $\nu=1$ bosonic Moore-Read Pfaffian 
\cite{MOORE1991,Greiter1991,Regnault2003} supporting non-Abelian topological order, can also be 
grown using our technique. Moreover, preparing bosons in higher LLs opens the 
possibility to simulate exotic Haldane pseudo-potentials, mimicking the effect of 
long-range interactions without the need to implement these in first place. We also expect that our scheme can be
adapted for the preparation of fractional quantum Hall states of fermions.

\paragraph*{Acknowledgements}
The authors thank N.Yao, M. Lukin, M. H\"oning and N. Lauk for helpful discussions. Support was provided by the NSF-funded Physics Frontier Center at the JQI and by ARO MURI Grant No. W911NF0910406. F.G. received support through the Excellence Initiative (DFG/GSC 266) and he gratefully acknowledges financial support from the "Marion K\"oser Stiftung". Financial support by the DFG within the SFB/TR 49 is also acknowledged.

\appendix

\section*{Supplementary: Laughlin states on the buckyball lattice}
To simulate fractional Chern insulators -- i.e. the lattice analogues of Laughlin states -- on a finite lattice system without edges, we consider bosons
hopping on the bonds between the 60 sites of a buckyball. All hopping elements are assumed to be
of magnitude $J$ and their phases are chosen in such a way that in total an integer amount $N_\phi$
of flux quanta pierce the surface, with homogeneous flux per plaquette $\alpha$ (in units of the flux quantum). This is a simple generalization of the
sphere surrounding a magnetic monopole which was introduced by Haldane \cite{Haldane1983}.

\begin{figure}[t!]
\centering
\epsfig{file=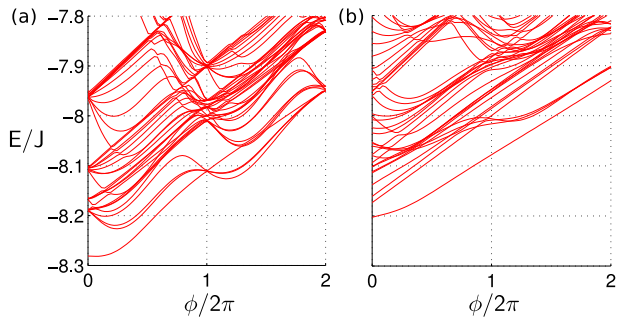, width=0.45\textwidth}
\caption{(Color online) Many-body spectra for $N=3$ bosons on a $C_{60}$ buckyball threaded by $N_\phi=4+\phi/2 \pi$ flux quanta. (a) Without trapping potential $g=0$, a gapped LN-type state is observed for $\phi=0$, which turns into a manifold of degenerate quasihole states for $\phi \geq 2 \pi$. (b) When a weak trapping potential Eq.\eqref{eq:Hpot} is switched on (here $g=J$), the ground state is gapped for all $\phi$. $U=10 J$ was used in both cases and for $\phi=0$ ($\phi=4 \pi$) we have $\alpha=0.125$ ($\alpha=0.1875$).}
\label{fig:C40spectra}
\end{figure}

Because in the spherical geometry -- unlike in the case of a torus -- Chern numbers can not readily be calculated from geometric Berry phases, we need to chose an alternative approach to identify Laughlin (LN) type ground states. To this end we adiabatically introduce magnetic flux through a single plaquette (say at the north pole), thereby increasing the charge of the fictitious magnetic monopole in the center of the sphere. This corresponds to the flux insertion described in the main text. As a consequence, an outwards Hall current pointing from north to south pole is generated, which is proportional to the Chern number of the many body state.

In FIG.\ref{fig:C40spectra} we show the flux-insertion spectra (i.e. the 
eigenenergies as a function of $\phi$) for $N=3$ bosons on the buckyball lattice. In (a) we did not
include the trapping potential Eq.\eqref{eq:Hpot} from the main text, and thus for $N_\phi=4$ we expect an incompressible LN-type ground state from the 
condition $N_\phi=2 (N-1)$ for $\nu=1/2$ LN states on a sphere. Indeed, we observe a 
ground state gap of the order $\Delta_\LN \approx 0.1 J$ in (a) as predicted for a 
square lattice \cite{Sorensen2005,Hafezi2007}. Moreover, for $\phi = 2 \pi$ and $4 
\pi$ the correct counting of (nearly degenerate) quasihole states is obtained, 
supporting our assumption that the ground state is in the LN universality class.

In FIG. \ref{fig:C40spectra} (b) the trapping potential Eq.\eqref{eq:Hpot} from the main text is included and the quasihole degeneracy is split. For all values of the additional magnetic flux $\phi$ a gapped ground state is observed, and by calculating the corresponding density profiles we checked that in the flux insertion the expected Hall-current corresponding to a Chern number $C=1/2$ is generated. This moreover shows that our intuitive picture of the ground state -- consisting of a quasihole trapped by the potential -- applies. Finally, by adiabatically increasing $g$ from $g=0$ in (a) to $g=J$ in (b) we checked that the ground state gap does not close at $\phi=0$ and its topological properties are thus unchanged by the potential Eq.\eqref{eq:Hpot} from the main text.

\end{document}